\documentclass[aps, prd, superscriptaddress, nofootinbib, preprintnumbers,twocolumn, showpacs]{revtex4}
\usepackage{graphicx}
\usepackage{amsmath}
\def\fsl#1{\setbox0=\hbox{$#1$}                 
   \dimen0=\wd0                                 
   \setbox1=\hbox{/} \dimen1=\wd1               
   \ifdim\dimen0>\dimen1                        
      \rlap{\hbox to \dimen0{\hfil/\hfil}}      
      #1                                        
   \else                                        
      \rlap{\hbox to \dimen1{\hfil$#1$\hfil}}   
      /                                         
   \fi}                                         %
\newcommand{\diag}{\mbox{diag}}

\newcommand{\VEV}[1]{\langle #1 \rangle}
\preprint{UWO-TH-05/12}
\begin{document}
\title{Neutral Larkin--Ovchinnikov--Fulde--Ferrell state 
and chromomagnetic instability in two-flavor dense QCD}

\author{E.V. Gorbar}
  \email{egorbar@uwo.ca}
  \altaffiliation[On leave from ]{
       Bogolyubov Institute for Theoretical Physics,
       03143, Kiev, Ukraine}
\author{Michio Hashimoto}
  \email{mhashimo@uwo.ca}
\author{V.A. Miransky}
  \email{vmiransk@uwo.ca}
   \altaffiliation[On leave from ]{
       Bogolyubov Institute for Theoretical Physics,
       03143, Kiev, Ukraine}
\affiliation{
Department of Applied Mathematics, University of Western
Ontario, London, Ontario N6A 5B7, Canada
}

\date{\today}

\begin{abstract}
In two-flavor dense quark matter, we describe the dynamics  
in the single plane wave Larkin--Ovchinnikov--Fulde--Ferrell (LOFF)
state satisfying the color and electric neutrality conditions.
We find that because the neutral LOFF state itself suffers from a 
chromomagnetic 
instability in the whole region where it
coexists with the (gapped/gapless)
two-flavor superconducting (2SC/g2SC) phases, it cannot cure
this instability in those phases.
This is unlike the recently revealed gluonic phase 
which seems to be able to resolve this problem.

\end{abstract}


\maketitle

It is natural to expect that cold quark matter may exist in the interior of
compact stars. This fact motivated intensive studies of this system over the
past few years (for a review, see Ref. \cite{RW}).
Matter in the bulk of stars should be electrically neutral.
It should be also a color singlet.
The electric and color neutrality conditions play an important role
in the dynamics of quark pairing \cite{AR,HS1}.
Several states have been studied in two and three-flavor
dense QCD as possible ground states satisfying these
conditions \cite{AR,HS1,Alford:2003fq,Reddy:2004my,Ruster,Abuki}.

Recently, it has 
been revealed that the (gapped/gapless)
two-flavor superconducting (2SC/g2SC) phase suffers from a chromomagnetic
instability connected with the presence of imaginary (tachyonic)
Meissner masses of gluons~\cite{HS2}.
Later a chromomagnetic instability has been also found in the color-flavor
locked phase of three-flavor QCD \cite{CFL}.   
It is clear that the resolution of this problem is intimately connected 
with the determination of the genuine ground state in dense quark matter.  
Therefore it is one of the central issues in the field.

In this Letter, we will study this problem in neutral two-flavor
dense QCD. At present, there are two types of dynamics suggested for its 
resolving. In Refs. \cite{GIR,H,Hong:2005jv}, the single plane
Larkin--Ovchinnikov--Fulde--Ferrell (LOFF)-like dynamics 
\cite{LOFF,LOFF-QCD} were
considered. [Although the dynamics considered in Refs. 
\cite{H,Hong:2005jv} look different from the conventional
LOFF one, by using an appropriate gauge
transformation, one can show that they are in fact equivalent to
the latter.]
The second type is the dynamics of the gluonic phase introduced and
studied recently in Ref. \cite{gluonic}. In that phase,
gluonic degrees of freedom play a crucial role and the dynamics
is manifestly non-abelian. It is clear that it would be 
important to discriminate these two types of dynamics in
order to establish the genuine ground state. It is the primary
goal of this Letter.

We recall that in the two-flavor case, the manifestations of the 
chromomagnetic instability are quite different in the regimes
with $\delta\mu < \Delta < \sqrt{2}\delta\mu$ and 
$\Delta < \delta\mu$ \cite{HS2}, where $\delta\mu$ yields   
a mismatch between chemical
potentials for the up and down quarks and $\Delta$ is
a diquark gap. The (strong coupling) regime with 
$\delta\mu < \Delta$  
corresponds to the 2SC solution, and the (intermediate coupling)
regime with $\Delta < \delta\mu$ corresponds to the gapless
g2SC one \cite{HS1}. While in the g2SC solution both
the 4-7-th gluons and the 8-th one have tachyonic masses, 
in the 2SC solution, with $\delta\mu < \Delta < \sqrt{2}\delta\mu$, only
the 4-7-th gluons are tachyonic. 

The main result of this Letter is that the neutral LOFF
state itself suffers from a
chromomagnetic
instability in the whole region where it
coexists with the 2SC/g2SC states. Because of that, it cannot cure
this instability in those states.
This is unlike the gluonic phase
which seems to be able to resolve this problem.

Studying the neutral LOFF state in two-flavor quark matter
is a nontrivial problem.
Up to now, such studies have been done only in the {\it weak}
coupling dynamical regime in which neither the neutral 2SC nor 
neutral g2SC
solutions exist \cite{nLOFF}. Here we will consider the neutral LOFF
phase both in the strong coupling and intermediate
coupling regimes that allows to
clarify the role of the LOFF dynamics for curing the chromomagnetic
instability. 
 
To study dense two-flavor quark matter in $\beta$-equilibrium, we consider
a phenomenological Nambu-Jona-Lasinio (NJL) model
whose Lagrangian 
is 
\begin{equation}
  {\cal L} = \bar{\psi}(i\fsl{\partial}+\hat{\mu}\gamma^0)\psi
  +G_\Delta \bigg[\,(\bar{\psi}^C i\varepsilon\epsilon^{\alpha}
\gamma_5 \psi) 
(\bar{\psi} i\varepsilon\epsilon^{\alpha}\gamma_5 
\psi^C)\,\bigg],
 \label{Lag}
\end{equation}
where $\varepsilon^{ij}$ and $\epsilon^{\alpha\beta\gamma}$ are
the antisymmetric tensors in the flavor and color spaces, respectively
(as usual, we neglect the current quark masses).
In $\beta$-equilibrium, the
elements of the diagonal chemical potential matrix $\hat \mu$ 
for up ($u$) and down ($d$) quarks
are $\mu_{ur} = \mu_{ug} = \bar{\mu} - \delta \mu,\,
\mu_{ub} = \bar{\mu} - \delta \mu - \mu_8$ and
$\mu_{dr} = \mu_{dg} = \bar{\mu} + \delta \mu,\,
\mu_{db} = \bar{\mu} + \delta \mu - \mu_8$, where
$\bar{\mu} \equiv \mu - \delta\mu/3 + \mu_8/3$ and
$\delta \mu \equiv \mu_e/2$.
Here the subscripts $r$, $g$, and $b$ correspond to red, green and
blue quark colors, 
$\mu$ is the quark chemical potential 
(the baryon chemical potential $\mu_B$ is given by $\mu_B \equiv 3\mu$),
$\mu_e$ is the chemical potential for the electric charge, and
$\mu_8$ is the color chemical potential.
The latter is connected with the vacuum expectation value
(VEV) of the time component of the 8-th gluon \cite{GR}.

By using the auxiliary field
$\Phi^{\alpha} 
\sim i\bar{\psi}^C\varepsilon \epsilon^{\alpha} \gamma_5 \psi$,
the Lagrangian density (\ref{Lag}) can be rewritten as
\begin{eqnarray}
  {\cal L} &=& \bar{\psi}(i\fsl{\partial}+\hat \mu \gamma^0)\psi
 - \frac{|\Phi^{\alpha}|^2}{4G_\Delta}  \nonumber \\
&&
 -\frac{1}{2}\Phi^{\alpha}
   [i\bar{\psi}\varepsilon\epsilon^{\alpha} \gamma_5 \psi^C]
 -\frac{1}{2}
   [i\bar{\psi}^C\varepsilon\epsilon^{\alpha}\gamma_5 \psi]\Phi^{*\alpha} .
 \label{Lag_aux}
\end{eqnarray}
The 2SC and g2SC states have a nonzero diquark condensate
$\VEV{\Phi^{\alpha}}|_{\alpha = b} \equiv \Delta$ chosen
along the blue color direction. As a result, the color group
$SU(3)_c$ is spontaneously broken to $SU(2)_c$ and
blue quarks are gapless.

The order parameter of a single plane wave LOFF state
has the form
$\VEV{\Phi^b(x)} = \Delta e^{2i\vec q \cdot \vec x}$\,
\cite{LOFF-QCD},
with a constant modulus $\Delta$ and a constant vector
$\vec{q}$ in the exponent.
Then it is
convenient to perform a gauge transformation
$\psi \to \psi' = exp(-2\sqrt{3}i\vec{q}\!\!\cdot\!\!\vec{x}T^8)\,\psi,$
where $T^8=\frac{1}{2\sqrt{3}}\,\diag(1,1,-2)$ is the eighth generator of
$SU_c(3)$ in the fundamental representation. This gauge transformation trades
the $x$-dependent phase of the LOFF order parameter for the
$x$-independent term 
$-2\sqrt{3}\,\bar{\psi}\vec{\gamma}\!\cdot\!\vec{q}T^8\psi$ in 
the kinetic term for quarks. Obviously, the equality
\begin{equation}
  \vec q = \frac{1}{2\sqrt{3}}\, g\VEV{\vec A^{(8)}}
\label{eight}
\end{equation}
takes place, i.e.,
the vector $\vec{q}$ is proportional to the VEV of a spatial
component of the eighth gluon.

To derive the free energy (effective potential) we need to calculate 
the fermion determinant.
For this purpose, it is convenient to introduce  the Nambu--Gor'kov spinor 
$\Psi^T \equiv
(\psi_{ir},\psi_{ir}^C,\psi_{ig},\psi_{ig}^C,\psi_{ib},\psi_{ib}^C)$.
Then, noting
that the contribution of the gapless blue quarks factorizes and applying 
the standard
formulas for 2 by 2 matrices in the red and green quarks sector, we find
\begin{widetext}
\begin{eqnarray}
\lefteqn{ \hspace*{-1.3cm}
  \det{\!}S^{-1}(p_0,\vec{p}) =
  \prod_{\tau^3=\pm 1}\bigg[(p^0+\bar{\mu}-\delta\mu\tau^3-\mu_8)^2
                        -(\vec{p}-2\vec{q})^2\bigg]^2
  \prod_{\tau^3=\pm 1}\bigg[(p^0-\bar{\mu}+\delta\mu\tau^3+\mu_8)^2
                        -(\vec{p}+2\vec{q})^2\bigg]^2
} \nonumber \\[-1mm] &&
  \times \prod_{\tau^3=\pm 1}\Bigg[\left\{
   \left(\,p^0-\delta\mu\tau^3 - Q\,\right)^2
    -(E^-_{\Delta,q})^2\right\}\,\left\{
   \left(\,p^0-\delta\mu\tau^3 +Q\,\right)^2
    -(E^+_{\Delta,q})^2\right\} + 4\Delta^2(Q^2-q^2)\,\Bigg]^4,
\label{determ}
\end{eqnarray}
\end{widetext}
where we defined
\begin{eqnarray}
Q&=& \frac{1}{2}\,\bigg(|\vec{p}+\vec{q}|-|\vec{p}-\vec{q}|\bigg),\,\,
q \equiv |\vec{q}|\,, \\
E^\pm_{\Delta,q} &\equiv& \sqrt{\frac{1}{4}\left(\,
   E^\pm_{p+q}+E^\pm_{p-q}\,\right)^2+\Delta^2}\,, \\
  E^\pm_{p+q} &\equiv& |\vec{p}+\vec{q}| \pm \bar{\mu}\,.
\end{eqnarray}

The free energy is expressed through an integral of
$\ln\det{\!}S^{-1}$. Because the  
vector $\vec{q}$ in the LOFF phase is just a 
redundant variable for gapless blue quarks, one can put 
$\vec{q} = 0 $ in the part of the free energy connected with
these quarks. 
Then it is obvious that their contribution to the free energy
coincides with that of free massless fermions.
In order to perform the loop integral for the red and green quarks part,
we make a usual assumption of the dominance of the region around
the Fermi surface, with $p\equiv |\vec{p}| \sim \bar{\mu}$. Then,  
neglecting contributions suppressed by  $1/\bar{\mu}^2$, we can omit
the last term $4\Delta^2(Q^2-q^2)$ in Eq. (\ref{determ}) and use the 
approximate relations
\begin{equation}
  Q= q\cos\theta + {\cal O}\left(\frac{q^3}{p^2}\right),
  \qquad \cos\theta \equiv \frac{\vec{p}\cdot\vec{q}}{p\,q},
  \label{app1}
\end{equation}
\begin{equation}
  E^\pm_{\Delta,q} = E^\pm_\Delta
  +\frac{q^2}{2p}\frac{p \pm \bar{\mu}}{E^\pm_\Delta}(1-\cos^2\theta)
  + {\cal O}\left(\frac{q^4}{p^2E^\pm_\Delta}\right),
  \label{app2}
\end{equation}
where 
$E^\pm_\Delta \equiv \sqrt{(p\pm\bar{\mu})^2 + \Delta^2}$.
Evaluating now the loop integral in the hard dense loop 
approximation, we 
obtain the following expression for the free energy of
the two-flavor LOFF state:
\begin{widetext}
\begin{equation}
  \Omega(\bar{\mu},\delta\mu,\mu_8,\Delta,q) =
  \Omega_{\rm 2SC} +\frac{2\bar{\mu}^2q^2}{3\pi^2} +
   \frac{\bar{\mu}^2}{\pi^2}\bigg[\,\frac{(q+\delta\mu)^3}{q}\bigg(\,
   \frac{1}{2}\,(1-x_1^2)\,\ln \frac{1+x_1}{1-x_1}
    - x_1 + \frac{2}{3}x_1^3\,\bigg) + (q \to -q)\,\bigg],
\label{V}
\end{equation}
with the 2SC part $\Omega_{\rm 2SC}$ being
\begin{equation}
  \Omega_{\rm 2SC} =
  - \frac{\mu_e^4}{12\pi^2}
  - \frac{\mu_{ub}^4}{12\pi^2} - \frac{\mu_{db}^4}{12\pi^2}
  - \frac{\bar{\mu}^4}{3\pi^2}
  + \frac{\Delta^2}{4G_\Delta} 
  - \frac{\bar{\mu}^2\Delta^2}{\pi^2}\,
    \ln \frac{4(\Lambda^2-\bar{\mu}^2)}{\Delta^2}
  - \frac{\Delta^2}{\pi^2}\left(\Lambda^2 - 2\bar{\mu}^2\right)\,,
\end{equation}
\end{widetext}
where $\Lambda$ is a ultraviolet cutoff in the NJL model,
$x_1 \equiv \theta(1-\frac{\Delta^2}{(\delta\mu+q)^2})\,
            \sqrt{1-\frac{\Delta^2}{(\delta\mu+q)^2}}$
with the step function $\theta(x)$, and
the contribution of electrons was also included in $\Omega_{\rm 2SC}$.

The following remarks are in order. Because we use the sharp cutoff
$\Lambda$ in evaluating the loop integral, it is necessary to subtract 
the corresponding vacuum contribution with 
$\bar{\mu}=\mu_8=\delta\mu=\Delta=0$ and the same value of the vector
$\vec{q}$: This leads to removing the spurious term 
$\Lambda^2\vec{q}{\:}^2$ in the free energy.
One can easily check that 
for
$q=0$, expression
(\ref{V}) yields the correct 2SC/g2SC free energy,
and for $\Delta = 0$, $q=0$, it reproduces the correct free energy for the
normal phase \cite{footnote}. Last but not least,
as will be discussed below, the one-loop
(i.e., mean field) approximation we use is reasonable for
realistic values of the chemical potential $\mu$.

The equations of motion for the free energy (\ref{V})
yield four equations:
Two gap equations for $\Delta$ and $q$ and two neutrality conditions
for $\delta\mu$ and $\mu_8$:
\begin{equation}
a)\,\,\frac{\partial \Omega}{\partial \Delta} = 0\,,\,\,\,\,\,
b)\,\,\frac{\partial \Omega}{\partial q} = 0\,,\,\,\,\,\,
c)\,\,\frac{\partial \Omega}{\partial \delta\mu} = 0\,,\,\,\,\,\,
d)\,\,\frac{\partial \Omega}{\partial \mu_8} = 0\,.
\label{gap}
\end{equation}
We analyzed both numerically and (in some limiting cases) 
analytically these equations. The analysis of
Eq. (\ref{gap}\,d) for 
$\mu_8$ shows that it is suppressed as
$\mu_8 \sim {\cal O}\left(\frac{\Delta^2}{\bar{\mu}}\right)$ or
$\mu_8 \sim {\cal O}\left(\frac{q^2}{\bar{\mu}}\right)$
that allows to put $\mu_8=0$ in Eqs. (\ref{V}), (\ref{gap}\,a), 
(\ref{gap}\,b), and (\ref{gap}\,c). 

\begin{figure}[tbp]
  \begin{center}
    \resizebox{0.42\textwidth}{!}{\includegraphics{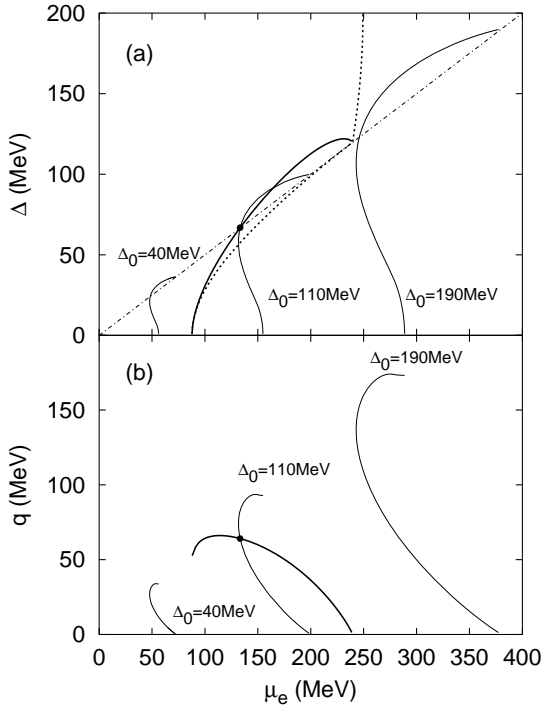}}
  \end{center}
\caption{The neutrality curve of the LOFF state (bold line)
         and solutions of the gap equations (thin solid lines)
         for three different values of $\Delta_0$.
         The dashed curve in (a) is the neutrality line for the 2SC/g2SC 
         state.
         The dash-dotted line in (a) is the $\Delta=\delta\mu$ line. 
\label{gapeq}}
\end{figure}

It is expected that the typical values of the quark chemical potential
$\mu$ in compact stars should be in the window $300-500$ MeV.
In the numerical analysis of equations (\ref{gap}),
we take $\mu$=400 MeV and $\Lambda$ = 653.3 MeV \cite{HS2}
[as will be discussed below, the results are not sensitive to the
choice of the value of $\mu$ in the interval $300-500$ MeV]. The 
NJL coupling
constant $G_{\Delta}$ determines physical properties of the model. In 
practice,
it is more convenient to trade $G_{\Delta}$ for $\Delta_0$, which is the
value of the 2SC gap at $\delta \mu$=0. Each of Eqs. (\ref{gap}\,a),
(\ref{gap}\,b), and (\ref{gap}\,c) with $\mu_8=0$ determines 
a surface in three dimensional space with coordinates $\Delta$, $q$, and 
$\mu_e = 2\delta\mu$. 
The intersection of the neutrality surface, determined by Eq. 
(\ref{gap}\,c),
and the surface determined by gap equation (\ref{gap}\,b) 
yields a neutrality curve. Its projections on the $(\mu_e,\Delta)$
and $(\mu_e,q)$ planes are shown in Figs. \ref{gapeq}\,(a) and 
\ref{gapeq}\,(b), 
respectively.
For a fixed $\Delta_0$, the solution for a neutral LOFF state is 
depicted
by a black bold point which is an intersection of the neutrality 
curve and 
the ``gap'' curve determined by two gap equations (\ref{gap}\,a) and 
(\ref{gap}\,b).  
In Fig.\ref{gapeq}, it is shown
the neutrality curve of the LOFF state (bold line) and three
gap curves (thin lines) for three characteristic
values of $\Delta_0$: 40 MeV (weak coupling), 110 MeV (intermediate 
coupling), and 190 MeV (strong coupling). The neutral LOFF
solution occurs only for the intermediate coupling, corresponding
to the gapless g2SC solution. 

The analysis of Eqs.(\ref{gap}\,a)-(\ref{gap}\,c) shows that the 
neutral LOFF
solution exists in the window
\begin{equation}
  \mbox{63 MeV} < \Delta_0 < \mbox{137 MeV} \quad  
  \mbox{(LOFF window)}\,,
\label{LOFFwindow}
\end{equation}
which can be compared with the g2SC window
\begin{equation}
 \mbox{92 MeV} < \Delta_0 < \mbox{130 MeV} \quad 
 \mbox{(g2SC window)}\,.
\label{g2SCwindow}
\end{equation}
We emphasize that for the same value $\Delta_0$ (i.e., the same value
of the coupling $G_{\Delta}$), the values of the parameters
$\Delta$ and $\delta\mu$ are different for the neutral 2SC/g2SC and
neutral LOFF phases.

\begin{figure}[tbp]
  \begin{center}
    \resizebox{0.47\textwidth}{!}{\includegraphics{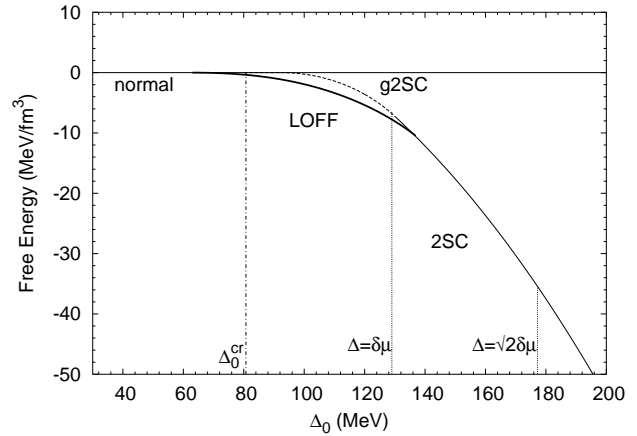}}
  \end{center}
\caption{Free energy of the neutral LOFF state (bold line),
        the neutral 2SC state (solid thin line), and the g2SC state 
        (dashed line). The free energy of the normal state 
        is chosen as a reference point. \label{energy_diff}}
\end{figure}

In the LOFF window (\ref{LOFFwindow}), we find that
$\Delta/\Delta_0=0\mbox{--}0.83$. 
The calculated free energies in different phases are shown in
Fig. \ref{energy_diff} (the free energy of the normal phase
was chosen as a reference point) \cite{footnote2}. 
One can see that the 
neutral LOFF phase is more
stable than the neutral
normal phase in the whole LOFF window (\ref{LOFFwindow}) where these
phases coexist. It is also
more stable than
the neutral g2SC/2SC phase in the whole g2SC window (\ref{g2SCwindow})
plus the narrow region $\mbox{130 MeV} < \Delta_0 < \mbox{136 MeV}$
in the 2SC phase near its edge with the g2SC one.
However, because the chromomagnetic instability in the 2SC phase
occurs at $\delta\mu=\Delta/\sqrt{2}$, which corresponds to 
$\Delta_0=$ 177 MeV, 
the neutral LOFF solution cannot cure this instability in a wide region,
with $\mbox{136 MeV} < \Delta_0 < \mbox{177 MeV}$, in that phase.

\begin{figure}[tbp]
  \begin{center}
    \resizebox{0.42\textwidth}{!}{\includegraphics{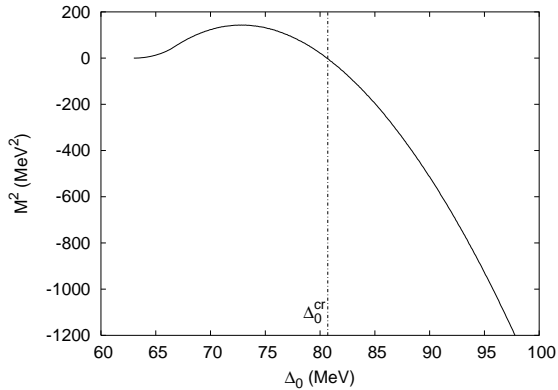}}
  \end{center}
\caption{Meissner mass square $M^2$ for the 4-7-th gluons
         in the neutral LOFF state.
         The QCD coupling constant $\alpha_s = g^2/(4\pi)$
         is chosen to be equal to 1.
\label{meissner}}
\end{figure}

It is however not the end of the story. Combining Eq. (\ref{gap})
with Eq. (68) for Meissner masses 
in the second paper in Ref. \cite{GIR}, we found that  
the Meissner masses for the 4--7-th gluons in neutral
LOFF state are imaginary for all values
of the parameter $\Delta_0 > \Delta_0^{\rm cr}=$ 81 MeV,
see Fig. \ref{meissner}
(the Meissner mass of the 8-th gluon is zero \cite{GIR}). 
{\it In other words,
for these values of $\Delta_0$, the neutral LOFF state
itself suffers from a chromomagnetic instability.}   
Because 
the values of $\Delta_0$ both in the 2SC phase and in
the g2SC window (\ref{g2SCwindow}) are larger than this critical
value $\Delta_0^{\rm cr}=$ 81 MeV (see Fig. \ref{energy_diff}), 
we conclude that in the whole
region where the neutral 2SC/g2SC and LOFF phases coexist, all
these three phases are unstable. {\it Therefore, the neutral LOFF state
with a single plane wave cannot cure the chromomagnetic instability.}   

We checked that this conclusion is not sensitive to the choice
of the value of the quark chemical potential in the
interval 300 MeV $< \mu <$ 500 MeV.
In particular, we found that for $\mu$= 300 MeV, the LOFF window is
$\mbox{49 MeV} < \Delta_0 <\mbox{121 MeV}$ and the g2SC window is
$\mbox{72 MeV}< \Delta_0 <\mbox{115 MeV}$. The neutral LOFF state has the
chromomagnetic instability for
$\Delta_0 > \Delta_0^{cr} = \mbox{64 MeV}$ in this case.
As to $\mu$= 500 MeV, the LOFF and g2SC windows are
$\mbox{74 MeV} < \Delta_0 < \mbox{141 MeV}$ and
$\mbox{109 MeV} < \Delta_0 < \mbox{134 MeV}$, respectively.
The neutral LOFF state
has the chromomagnetic instability for $\Delta_0 > \Delta_0^{cr}$ = 94 MeV.
Thus, the conclusions of the present analysis seem to be robust.

As to the reliability of the mean field (one-loop)
approximation we use in this problem, Fig. \ref{gapeq}\,(a) shows
that the values of the ratio $\Delta/\mu$ in the neutral LOFF
state are typically less than one fourth. This suggests that the mean
field approximation should be at least qualitatively
reliable. Still, it would be certainly important to clarify
more explicitly the role of fluctuations in this problem.

Because of the relation between the LOFF vector $\vec{q}$ and
the VEV of $\vec{A}^{(8)}$ in Eq.(\ref{eight}), the
chromomagnetic instability in the neutral LOFF state
implies that the vector condensate $\VEV{\vec{A}^{(8)}}$ 
alone is not enough for washing out the chromomagnetic instability.
We consider this as a strong indication on the relevance of
the gluonic phase \cite{gluonic} for curing this problem. Indeed,
in the gluonic phase there are additional vector condensates
directly connected with the 4--7-th gluons and, as was shown
in Ref. \cite{gluonic}, there is no chromomagnetic instability in
that phase in the strong coupling regime close to the value  
$\Delta_0 =$ 177 MeV corresponding to the end point 
$\delta\mu = \Delta/\sqrt{2}$ of this instability in the 2SC solution.  

This makes the study of
the gluonic phase  
both far away from the edge $\Delta_0 =$ 177 MeV
of the chromomagnetic instability 
in the 2SC solution and in the g2SC window to be quite desirable. 
Also, it is clear that
it would be interesting to study the neutral LOFF state in 
the three-flavor QCD \cite{Casalbuoni:2005zp} and to examine
the relevance of
the LOFF state with more than one plane waves \cite{BR}
for curing the chromomagnetic instability.

The work was supported by the Natural Sciences and Engineering Research
Council of Canada.

\begin{thebibliography}{99}
\bibitem{footnote}
While the $\vec{q}$ dependent part of the free energy (\ref{V})
coincides with that in the second paper in Ref. \cite{GIR},
the 2SC parts of these two expressions are essentially different.
As a result, the free energy in Ref. \cite{GIR}
does not reproduce correctly
the limits with $q=0$ and with $q=0$, $\Delta=0$. Because
of that, it 
is not appropriate for studying a neutral LOFF state.
\bibitem{footnote2}
In the part of LOFF window (\ref{LOFFwindow}) with 
130 MeV $<\Delta_0<$137 MeV, there exists a second branch
of the neutral LOFF solution. By using expression (\ref{V}),
one can show that its free energy is larger than the
free energy of the first branch.
This energetically unfavorable branch is not
shown in Fig. \ref{energy_diff}.
\end{thebibliography}
\end{document}